\documentclass[aps,prb,twocolumn,amsmath,amssymb,superscriptaddress,floatfix]{revtex4}
\usepackage{graphicx}
\usepackage{graphics}
\usepackage{bm}
\usepackage[usenames]{color}
\usepackage{colordvi}
\usepackage{units}
\usepackage{bbm}
\usepackage{booktabs}
\usepackage{array}
\usepackage{placeins}
\usepackage{epsfig}
\usepackage{lscape}
\usepackage{changes}
\usepackage{braket}
\usepackage{tabularx}
\newcolumntype{L}[1]{>{\raggedright\arraybackslash}p{#1}} 
\newcolumntype{C}[1]{>{\centering\arraybackslash}p{#1}} 
\newcolumntype{R}[1]{>{\raggedleft\arraybackslash}p{#1}} 
\usepackage[colorlinks=true,linkcolor=blue,citecolor=blue,urlcolor=blue]{hyperref}
\newcommand{\be}{\begin{equation}}
\newcommand{\ee}{\end{equation}}
\newcommand{\beqn}{\begin{eqnarray}}
\newcommand{\eeqn}{\end{eqnarray}}

\definecolor{mymagenta}{rgb}{1.0,0.0,1.0}
\definecolor{mycyan}{rgb}{0.0,1.0,1.0}
\definecolor{myyellow}{rgb}{1.0,1.0,0.0}
\definecolor{myorange}{rgb}{1.0,0.27,0.0}

\definecolor{dark-gray}{HTML}{a0a0a0}
\definecolor{dark-red}{HTML}{8b0000}
\definecolor{dark-green}{HTML}{006400}
\definecolor{dark-blue}{HTML}{00008b}
\definecolor{gold}{rgb}{1.0,0.84,0.0}
\definecolor{dark-turquoise}{HTML}{00ced1}

\bibstyle{apsrev.bib}

\begin{document}

\title{Geometry of rare regions behind Griffiths singularities in random quantum magnets}
\author{Istv\'an A. Kov\'acs}
\email{istvan.kovacs@northwestern.edu}
\affiliation{Department of Physics and Astronomy, Northwestern University, Evanston, IL 60208-3112, USA}
\affiliation{Northwestern Institute on Complex Systems, Northwestern University, Evanston, IL, USA}
\author{Ferenc Igl{\'o}i}
\email{igloi.ferenc@wigner.hu}
\affiliation{Wigner Research Centre for Physics, Institute for Solid State Physics and Optics, H-1525 Budapest, P.O. Box 49, Hungary}
\affiliation{Institute of Theoretical Physics, Szeged University, H-6720 Szeged, Hungary}
\date{\today}

\begin{abstract}
In many-body systems with quenched disorder, dynamical observables can be singular not only at the critical point, but in an extended region of the paramagnetic phase as well. These Griffiths singularities are due to rare regions, which are locally in the ordered phase and contribute to a large susceptibility. Here, we study the geometrical properties of rare regions in the transverse Ising model with dilution or with random couplings and transverse fields. In diluted models, the rare regions are percolation clusters, while in random models the ground state consists of a set of spin clusters, which are calculated by the strong disorder renormalization method. We consider the so called energy cluster, which has the smallest excitation energy and calculate its mass and linear extension in one-, two- and three-dimensions. Both average quantities are found to grow logarithmically with the linear size of the sample. Consequently, the rare regions are not compact: for the diluted model they are isotropic and tree-like, while for the random model they are quasi-one-dimensional.
\end{abstract}


\maketitle

\section{Introduction}
\label{sec:intr}

Randomness is an inevitable feature of real materials, and even a small amount of quenched disorder can alter the collective properties of interacting many-body systems. This is known at a phase transition point where the critical exponents could be modified at a second order transition point\cite{harris} or a first order transition could be smoothed to a continuous one due to randomness\cite{aizeman,cardy99}. Even in the paramagnetic phase, the presence of quenched disorder can result in unusual dynamical properties, which do not exist in pure systems. For example, in a random ferromagnet the linear susceptibility, $\chi$ can be divergent in an extended part of this region, which is called the Griffiths-phase\cite{griffiths}. This type of singular behaviour is due to rare regions\cite{vojta}, in which there are extreme fluctuations of strong couplings and domains are formed, which can remain locally ordered even in the paramagnetic phase. The relaxation time, $\tau$, associated with turning the spins in such domains can be extremely large and it has no upper limit in the thermodynamic limit. This type of Griffiths singularities are responsible for non-analytic behaviour of several average physical quantities, besides the susceptibility one can mention the specific heat or the auto-correlation function.

Randomness in quantum systems has stronger impact than in classical ones\cite{rieger_young97,bhatt}. The critical behaviour is often controlled by a so called infinite disorder fixed point\cite{daniel_review}, at which the linear length of the system, $L$, and the time-scale, $\tau$, is related as:
\be
\ln \tau \sim L^{\psi}\;,
\label{psi}
\ee
with $\psi$ being a critical exponent. In the Griffiths-phase the auto-correlation function has a power-law decay, $G(t) \sim t^{-d/z}$, and the relation between length-scale and time-scale is in the form:
\be
\tau \sim \varepsilon^{-1} \sim L^z\;.
\label{L^z}
\ee
Here, $\varepsilon$ denotes the excitation energy, $d$ is the dimension of the system and $z=z(\delta)$ is the dynamical exponent, which is a continuous function of the distance from the critical point, $\delta$. By approaching an infinite disorder critical point $z(\delta)$ is divergent: $d/z \sim \delta^{\nu \psi}$, $\nu$ being the critical exponent of the average correlation length: $\xi \sim |\delta|^{-\nu}$. In the Griffiths-phase the average susceptibility, $\chi$ and that of the specific heat, $c_V$, show power-law singularities at low temperatures, $T$:
\be
\chi(T) \sim T^{-1+d/z},\quad c_V(T) \sim T^{d/z}\;,
\ee
thus at zero temperature $\chi(0)$ is divergent for $z>d$.\cite{mccoy,vojta}.

In random quantum systems with discrete symmetry, the probability of small gaps can be derived from a scaling theory based on the assumption that the low-energy excitations are localized to rare regions\cite{thill_huse,young_rieger}. Then, the probability of having a small excitation energy is proportional to the volume of the system, $L^d$. Hence, the probability to have a small gap between $\varepsilon$ and $\varepsilon(1+\Delta)$ has the scaling form: $\Delta L^d \varepsilon^{d/z}$, which follows from Eq.(\ref{L^z}). Thus the cumulative distribution of small gaps is given by:
\be
\mu(\varepsilon)=1-A\epsilon^{d/z},\quad \varepsilon \ll 1\;,
\label{cumulated}
\ee
where $A>0$ is some constant.

Since the rare regions are well separated, one might assume that the related low-energy excitations are uncorrelated. In this case, the probability distribution function of the smallest gap in a finite system can be obtained from extreme statistics of independent and identically distributed (iid) random numbers having a parent distribution in Eq.(\ref{cumulated}). The result is the Fr\'echet distribution\cite{fisher_tippett,gumbel,galambos}:
\be
P_L(\varepsilon)=\frac{d}{z}u^{d/z-1}\exp\left(-u^{d/z}\right)\;,
\label{P_L_SDRG}
\ee
in terms of $u=u_0 \varepsilon L^z$ and $u_0$ is a constant. The validity of this relation has been studied numerically for the random transverse Ising model and presumably exact agreement is found asymptotically in the $L \to \infty$ limit in $d=1$\cite{fisher_young,jli}. The numerically found agreement is also satisfactory for $d=2$\cite{jli}. More recently, we have studied the finite size corrections to Eq.(\ref{P_L_SDRG}) for the $d=1$ system\cite{kpi} and systematic deviations are observed from the analytical results for iid random numbers. This fact shows that the weak correlations between low-energy excitations in random quantum magnets are relevant.

In the previous paragraphs we showed that the concept of rare regions for localized excitations is very useful in understanding the singular properties of Griffiths-phases in random quantum magnets. Here, we aim to visualize and understand the geometric characteristics of these rare regions.
Specifically, we want to address the question whether the monotonous decrease of the excitation energy with $L$ (see in Eq.(\ref{L^z})) goes hand in hand with the growth of these regions, or simply due to the appearance of stronger bonds in the same volume. In this paper, we carry out such an investigation on a paradigmatic system, the transverse Ising model (TIM), both with dilution (diluted model) and with random couplings and/or transverse fields (random model). For the diluted model, the low-energy clusters are the same as for percolation and we analyse the extreme statistics of these clusters. For the random model, the calculation is more involved and the Strong Disorder Renormalization Group (SDRG) method\cite{im} is used, in which the ground state of a given sample is represented by isolated sites and clusters of sites. To each cluster (and to each isolated site) an excitation energy is associated and the smallest one represents the lowest gap of the sample. The cluster with the smallest excitation energy (the so-called energy cluster) represents the rare region we are looking for. 

In the actual calculation, we used a numerical implementation of the SDRG method\cite{ddRG} and for a large number of realizations we measured the excitation energy and different geometrical properties of the energy cluster (number of sites, linear extension, radius of gyration) and studied the dependence of their average value on the linear size of the system. We have most detailed calculations in $d=1$ in which case we went up to $L=2^{15}$ and used $10^6$ realizations. In $d=2$ and $3$ the largest systems were $L=2^{10}$ and $2^8$, respectively, and we have at least $10^4$ realizations for the largest sizes.

The rest of the paper is constructed as follows. In Sec. \ref{sec:model} the diluted and the random TIMs are defined and the basic elements of the SDRG method are recapitulated. In Sec. \ref{sec:results} first the shape of the energy clusters in different dimensions are presented and their measured properties are defined. Afterwards, the measured average parameters of the energy clusters are shown and their size dependency is analyzed. Our paper concludes with a discussion in the final section.

\section{Models and the SDRG procedure}
\label{sec:model}

Here, we consider the transverse Ising model in a $d$-dimensional hypercubic lattice and the Hamiltonian is defined by
\be
\hat{H}=-\sum_{\langle i,j \rangle} J_{ij} \sigma_{i}^{x} \sigma_{j}^{x}
-\sum_{i} h_i \sigma^z_{i}\;,
\label{Hamilton}
\ee
in terms of the $\sigma_{i}^{x,z}$ Pauli matrices at site $i$ and the first sum runs over nearest neighbours.

\subsubsection{Diluted model}

\begin{figure}[h!]
\begin{center}
\includegraphics[width=1.0\linewidth]{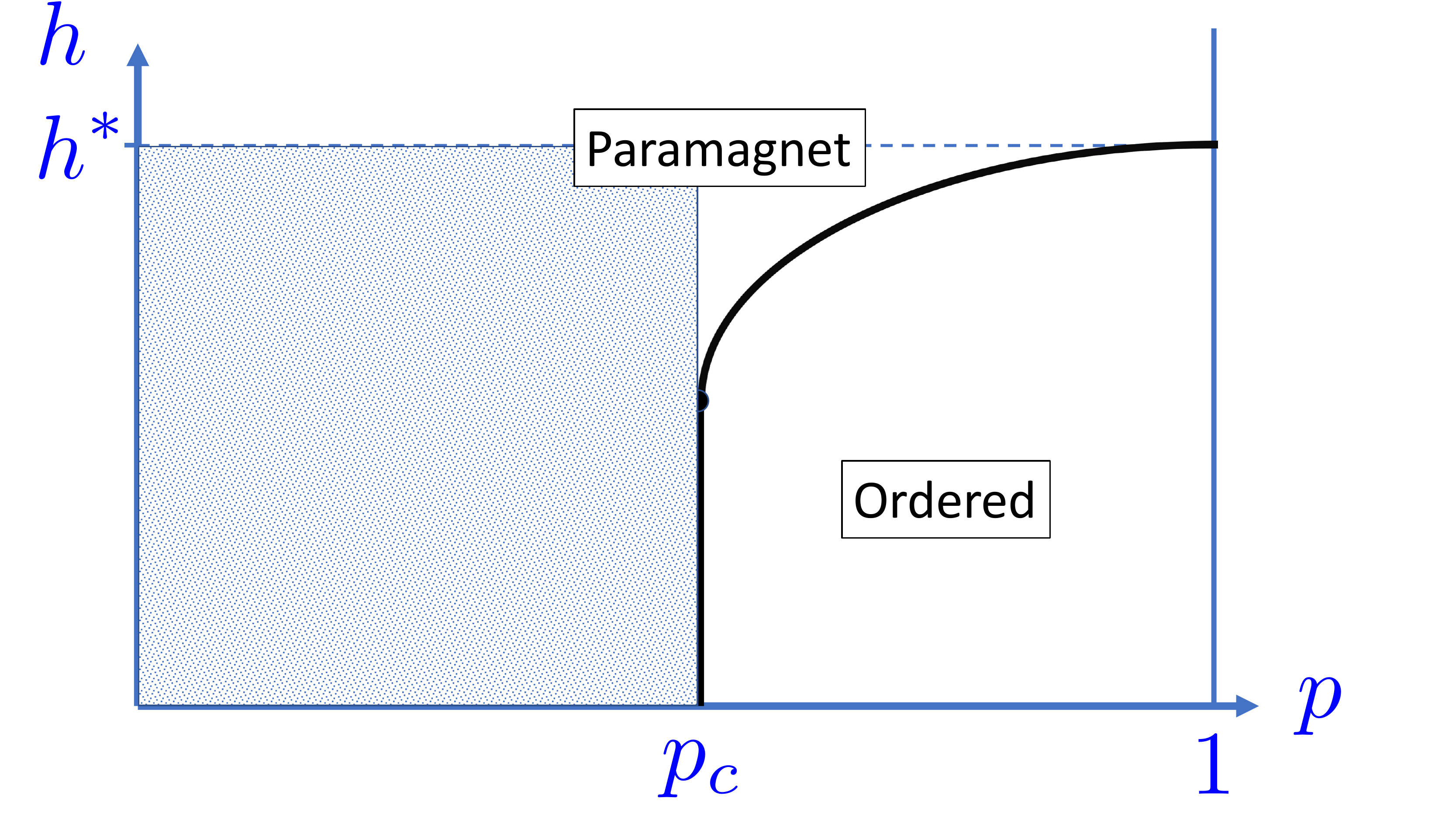}
\end{center}
\caption{\label{fig:diluted}(Color online) Schematic phase-diagram of the diluted TIM in dimensions $d \ge 2$. In this paper, the shaded region of the Griffiths-phase is considered. In one dimension $p_c=1$ and $h^*=1$.}
\end{figure}

In the bond diluted model, all the $\{h_i\}$ are equal to $h>0$ and the couplings are equal to $J=1$ with probability $p$ or equal to $J=0$ with probability $1-p$. We also consider the site diluted model, in which $J_{ij}=\epsilon_i \epsilon_j$ and $h_i=h \epsilon_i$ and $\epsilon_i=1$ with probability $p$ and $\epsilon_i=0$ with probability $1-p$. In 1D, for $p<1$, the system is in the paramagnetic phase: it is separated into finite segments. In higher dimensions, there is a percolation transition at $p=p_c$ and in the subcritical phase, $\delta=\Delta p=p-p_c<0$, the system consists of isolated clusters\cite{senthil_sachdev}. In this subcritical region, having a sufficiently strong coupling, such that the pure system is in the ferromagnetic phase, $h < h^*$ the finite clusters are trying to order, which results in a large susceptibility, a hallmark of the Griffiths-phase. We are going to study the system in this region, see in Fig.\ref{fig:diluted}.

\subsubsection{Random model}

In the random model, the couplings are taken from the uniform distribution:
\be
p(J)=
\begin{cases}
1,~{\rm for}~0 \le J \le 1\\
0,~{\rm otherwise}\;,
\end{cases}
\ee
and similarly for the transverse fields:
\be
q(h)=
\begin{cases}
h_0^{-1},~{\rm for}~0 \le h \le h_0\\
0,~{\rm otherwise}\;.
\end{cases}
\ee

\begin{figure}[h!]
\begin{center}
\includegraphics[width=1.0\linewidth]{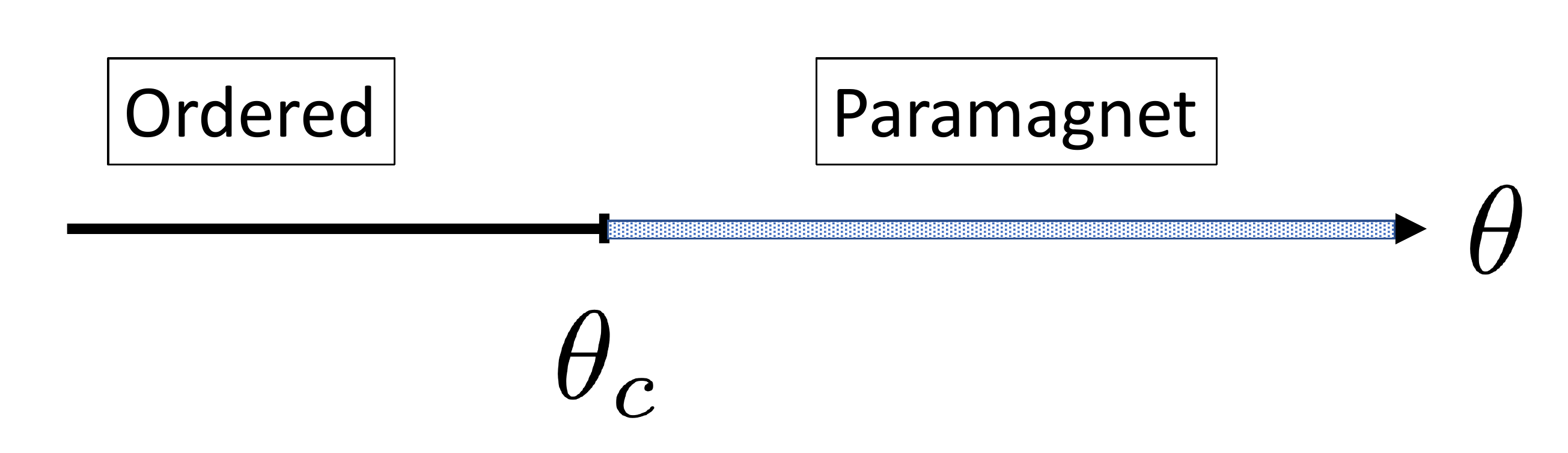}
\end{center}
\caption{\label{fig:random}(Color online) Schematic phase-diagram of the random TIM. The shaded region represents the Griffiths-phase, which extends to the complete paramagnetic phase.}
\end{figure}

Generally, the logarithmic variable, $\theta=\ln(h_0)$, is used as the control-parameter. The system is in the ferromagnetic and the paramagnetic phase for $\delta=\Delta \theta=\theta-\theta_c<0$ and $\Delta \theta>0$, respectively and the properties of the system at the critical point, $\theta_c$, are controlled by infinite disorder fixed points, see in Refs.\cite{fisher,2d,2dki,ddRG}. Since the minimal value of the transverse field is $h_{min}=0$, the Griffiths-phase extends to the complete paramagnetic phase, $\Delta \theta>0$, see in Fig.\ref{fig:random}.

\subsection{SDRG procedure}

We are going to study the properties of the random TIM by the SDRG procedure\cite{im}, which has been introduced by Ma \textit{et al.}\cite{mdh} and further developed by Fisher\cite{fisher}. The method provides analytical results in $d=1$\cite{fisher,Igloi} and accurate numerical ones in higher dimensions\cite{2d,2dki,ddRG}. In this procedure at each step of the renormalization the largest parameter in the Hamiltonian (either coupling or transverse field) is eliminated and new  terms are generated between remaining sites through second-order perturbation calculation. If a strong coupling, say $J_{ij}$, is decimated the two connected spins form a spin cluster having an additive moment, $\tilde{\mu}=\mu_i+\mu_j$. (At the starting step all spins have the same moment $\mu_i=1, \forall i$.) The spin cluster is placed in an effective transverse field of strength: $\tilde{h}=h_i h_j/J_{ij}$, which is obtained from the first gap of the cluster. If a large transverse field, say $h_i$, is decimated, the actual spin is eliminated and new effective couplings are generated between each pair of spins, which are nearest neighbours to $i$, say $j$ and $k$, having a value: $\tilde{J}_{jk}=J_{ji}J_{ik}/h_i$.

In $d=1$ the system remains a chain with fewer sites under decimation. In higher dimensions, however, the topology of the system is gradually changing. In this case it can happen, that at some step two parallel couplings appear between two neighboring sites, in which case the maximum of them is taken. The use of this "maximum rule" is asymptotically exact at an infinite disorder fixed point and results in simplifications in the renormalization algorithm. Using the optimized algorithm described in Refs.\cite{ddRG} the time to renormalize a cluster with $N$ sites and $E$ edges was $t \sim {\cal O}(N \ln N + E)$, which is to be compared with the performance of a na\"{\i}ve implementation: $t \sim {\cal O}(N^3)$.

The SDRG method is expected to provide asymptotically exact results at an infinite disorder fixed point, which has been checked in $d=1$ by comparing the results with those of analytical and numerical calculations\cite{young_rieger,bigpaper}. Also in $d=2$ the SDRG results (obtained by the use of the maximum rule) are consistent with quantum Monte Carlo simulations\cite{pich,matoz}. Regarding the Griffiths-phase, the SDRG describes the Griffiths singularities asymptotically exactly in $d=1$, where the dynamical exponent is given by the positive root of the equation~\cite{Z,ijl01,Igloi}:
\be
      \left[\left(\frac{J}{h} \right)^{1/z}\right]_{\rm av}=1\;.
      \label{eq:z_eq}
\ee
For the uniform distribution the dynamical exponent is given by:
\be
(1-z^{-2})h_0^{1/z}=1\;,
\label{eq:z_box}
\ee
while for the bond-diluted chain we have:
\be
z=\frac{\ln h}{\ln p}
\label{eq:z_diluted}
\ee
in agreement with the result of the calculation in Ref.\cite{daniel_review}.

The distribution of the smallest energy excitations obtained by the SDRG also agrees well with the precise numerical calculations, even for large but finite $L$ values\cite{kpi,alcaraz}. Therefore, we expect that the SDRG method provides valuable information about the statistics of energy clusters, too. 

We mention that in $d=1$, the statistics of clusters has already been studied at the critical point\cite{kkaf}. In this case, the linear extension of the largest cluster scales with the length of the chains, thus density profiles are numerically calculated, which - close to the open boundaries - are found to follow scaling predictions. Critical energy clusters have also been studied at $d=2$ and $d=3$, which are shown to span the finite system and have a quasi-one-dimensional structure\cite{ddRG}. In the Griffiths-phase, where dominantly transverse fields are decimated, the typical clusters have a finite extent, which is proportional to the correlation length. We are going to study the geometrical properties of the largest one, the energy cluster in the following sections.

\section{Results}
\label{sec:results}

\subsection{Typical rare regions}
\label{sec:typical_rare_regions}
 
\begin{figure}[h!]
\begin{center}
\includegraphics[height=0.7\linewidth]{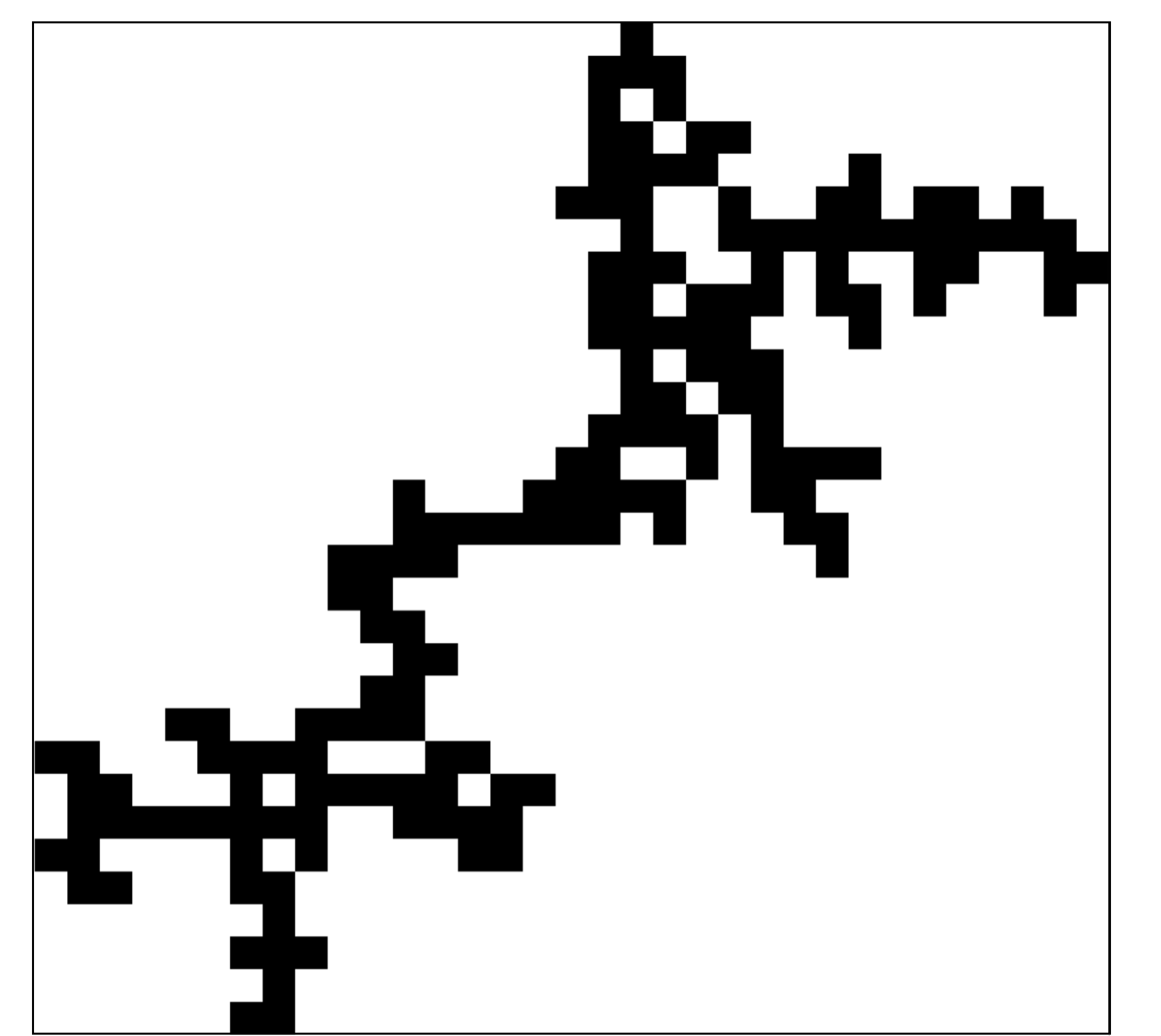}
\end{center}
\vskip-4mm
\caption{\label{fig:perc}(Color online) Typical 2D energy cluster of the site-diluted TIM with an outer frame at $\Delta p=-0.2$ for squares with $L=8,192$. The cluster is connected and has a tree-like structure.}
\end{figure}

The structure of the diluted model in 1D is trivial, it consists of separated connected segments, among which the energy cluster has the largest mass, which equals to its length. In higher dimensions, the ground state of the system is given by a set of percolation clusters, among which  the one with the largest mass is the energy cluster. A typical energy cluster for the site-diluted TIM is shown in Fig.\ref{fig:perc}.

\begin{figure}[h!]
\begin{center}
\includegraphics[width=1.0\linewidth]{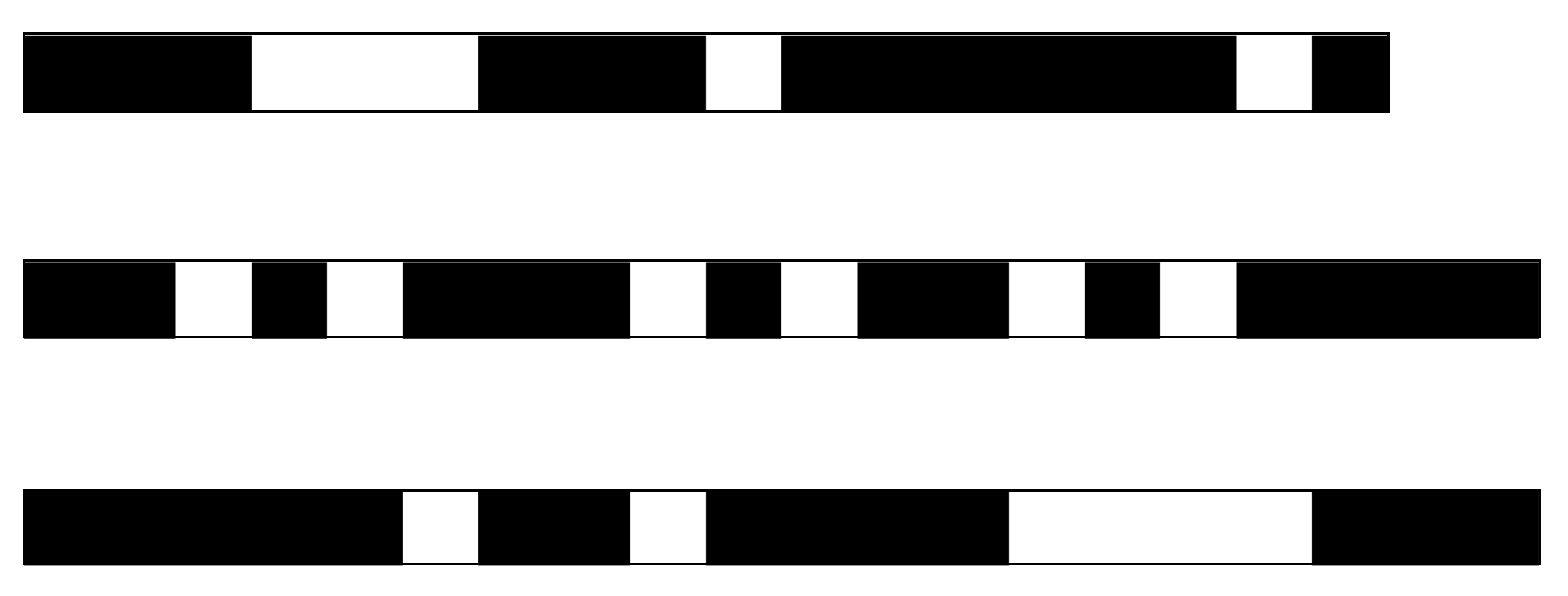}
\end{center}
\vskip-2mm
\caption{\label{fig:cluster1}(Color online) Typical 1D energy clusters at $\Delta \theta=\ln 2$ for chains with $L=32768$. A unit square is assigned to each site: black (white) squares denote non-decimated (decimated) spins. The clusters are $\ell=18,~20,~20$ in length and $\mu=13,~14,~14$ in mass from top to bottom.}
\end{figure}

\begin{figure}[h!]
\begin{center}
\includegraphics[width=1.0\linewidth]{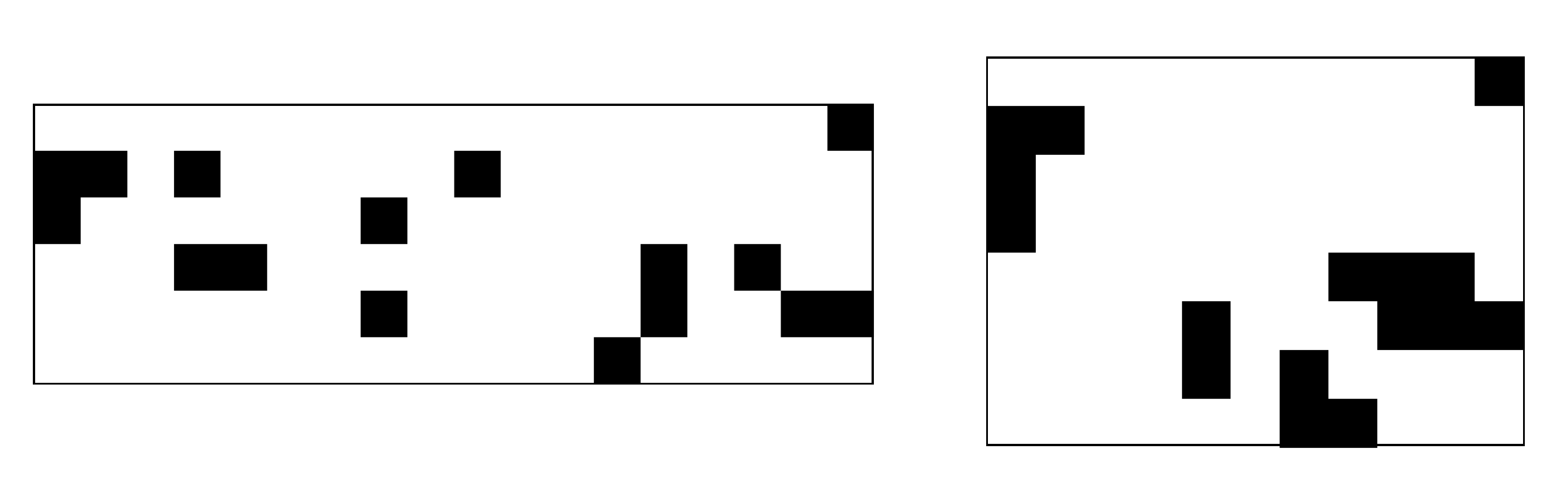}
\end{center}
\vskip-4mm
\caption{\label{fig:cluster2}(Color online) Typical 2D energy clusters of the random TIM with an outer frame at $\Delta \theta=0.5$ for squares with $L=4,096$. A unit square is assigned to each site: black (white) squares denote non-decimated (decimated) spins. The left (right) cluster have a linear extension (longer edge of the frame) $\ell=18~(11)$ and a mass $\mu=16~(16)$.}
\end{figure}

For illustration in the random model, a few energy clusters are shown in Fig.\ref{fig:cluster1} for $d=1$ and in Fig.\ref{fig:cluster2} for $d=2$. The energy clusters are represented by the contributing original (non-decimated) sites, which are indicated by black squares in the figures. Note that the energy cluster in the random model generally consists of several disconnected parts.

From a geometrical point of view, the energy clusters are characterised by the linear extension, $\ell$, the radius of gyration, $R_g$, and the mass, $\mu$. In $d=1$, $\ell$ is simply the length of the cluster. In higher dimensions, we draw an outer frame to the cluster (see in Figs.\ref{fig:cluster2} and \ref{fig:perc}) and the longest edge of this frame defines $\ell$. The mass of the cluster is the number of its sites, $N$, and the radius of gyration is defined as
\be
R_g^2=\frac{1}{N(N-1)} \sum_{i \ne j} [\vec{r_i}-\vec{r_j}]^2\;,
\ee
where $\vec{r_i}$, $i=1,2,\dots,N$ is the position of a site in the cluster.

\subsection{Numerical analysis}

For the diluted TIM, numerical calculations are performed in dimensions $d=2$ and $3$, for different linear sizes of the samples. We set the control parameter for a given value, $p<p_c$, (for site percolation $p_c=0.592746$ ($d=2$)\cite{blote,jacobsen} and $p_c=0.3116$ ($d=3$)\cite{ziff}), and determined the largest connected percolation cluster in the different samples.

For the random TIM, the numerical calculations are performed in $d=1,2$ and $3$. First, for a given dimension, we set the value of the control parameter in the Griffiths-phase for $\theta > \theta_{c}$. For the uniform distribution, the critical point is known exactly in $d=1$, $\theta_{c}=0$, whereas in $d=2$ and $d=3$ these were obtained before as $\theta_{c}=1.6784$ and $\theta_{c}=2.5305$, respectively, by the numerical application of the SDRG approach with the maximum rule\cite{ddRG}. Having a fixed value for $\theta$, we considered a large set of random samples of different finite length, $L$, and performed the renormalization procedure up to the last site. In a given sample, during the decimation process, spin clusters are formed and then eliminated. In this procedure, we have selected the energy cluster, which has the smallest excitation energy.

\subsection{Analysis of the 1D results}

\subsubsection{Diluted model}

In 1D, the critical point is at $p_c=1$ and the probability to have a cluster of size $\mu$ is given by $P_{\rm{1D}}(\mu)=p^{\mu}$. The typical value of the largest cluster $\mu_{\mathrm{typ}}$ in a system of length $L$ is given by the condition: $p^{\mu_{\mathrm{typ}}}L \approx 1$, thus 
\be
\mu_{\mathrm{typ}}(L) \approx \frac{\ln L}{|\ln p|} + {\rm cst.}\;.
\label{1d_diluted}
\ee
Using the estimate of the energy gap, $\epsilon=h^{\mu}\sim L^{-z}$, together with the value of the dynamical exponent in Eq.(\ref{eq:z_diluted}), we arrive at the result in Eq.(\ref{1d_diluted}).
The average value of the mass, as well as the average value of the length of the largest cluster, should scale as $\ln L$.

\subsubsection{Random model}

Here, we set the control parameter to $\theta=\ln(h_0)=\ln(2.)$, and renormalized finite systems of lengths: $L=16,32,64,\dots,32768$. For each length, we considered $10^6$ random samples and calculated the average value of the geometrical properties: $[\ell]_{\rm av}(L)$, $[\mu]_{\rm av}(L)$ and $[R_g]_{\rm av}(L)$, as well as the average value of the log-gap: $[\ln(\varepsilon)]_{\rm av}(L)$. (Here and in the following we use $[\dots]_{\rm av}$ to denote averaging over quenched disorder.) 

\begin{figure}[h!]
\begin{center}
\includegraphics[width=1. \linewidth]{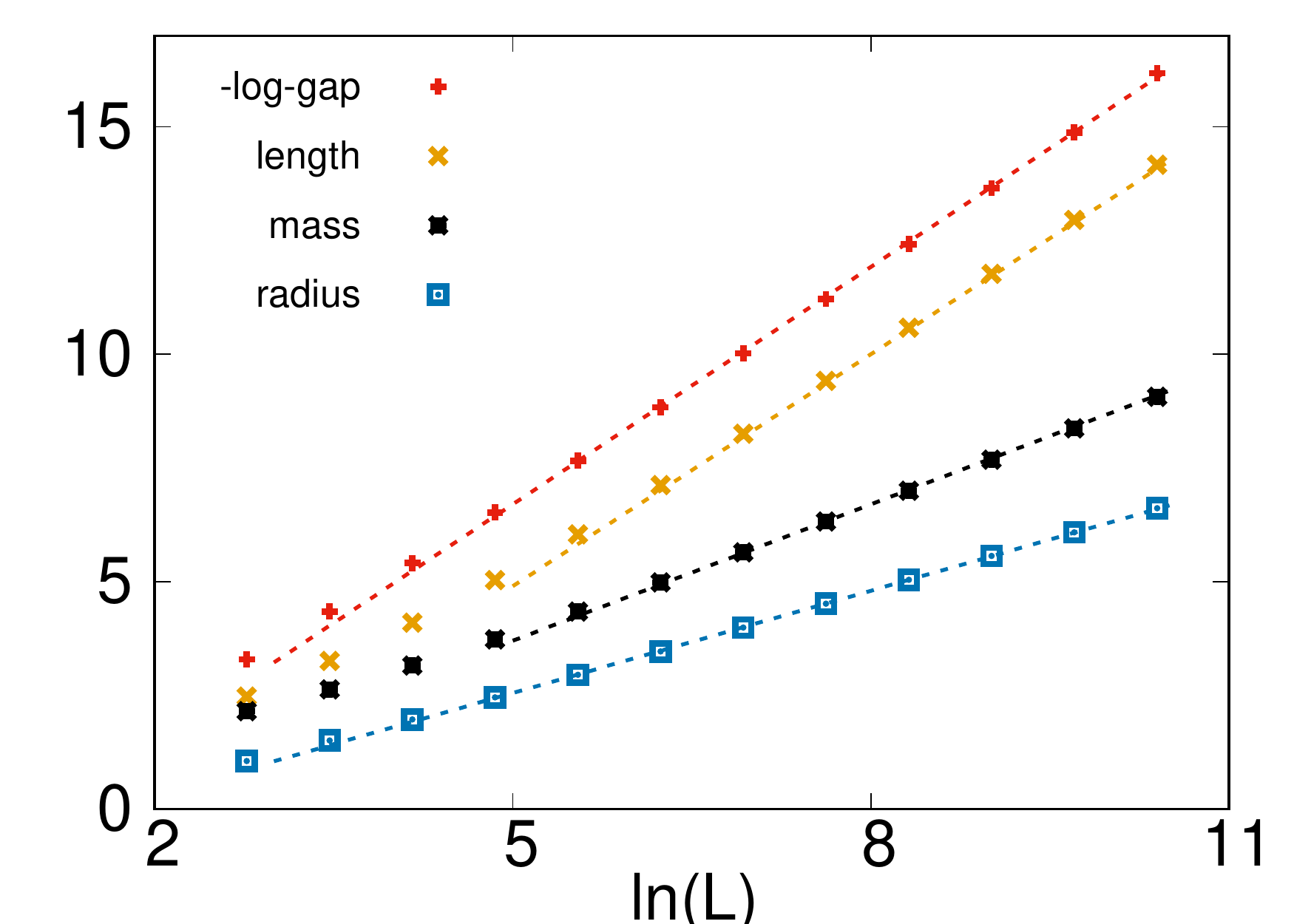}
\end{center}
\vskip-2mm
\caption{\label{fig:1D_parameters}(Color online) Average parameters of energy clusters in the 1D RTIM at $\theta=\ln 2$ as a function of the logarithm of the size of the system. From top to bottom: the minus log-gap $-[\ln(\varepsilon)]_{\rm av}$, the length $[\ell]_{\rm av}$, the mass $[\mu]_{\rm av}$ and the radius of gyration $[R_g]_{\rm av}$. The slopes of the straight lines are 1.74, 1.7, 1.0 and 0.75, from top to bottom.}
\end{figure}

The possible $L$-dependence of the cluster parameters is similar to that of the diluted model and can be obtained from the following consideration. The log-gap scales as 
\be
[\ln(\varepsilon)]_{\rm av}(L) \simeq z \ln L + {\rm cst.}\;.
\ee
which follows from Eq.(\ref{L^z}).
Within the SDRG procedure, this gap is given in terms of the transverse fields, $h_i$, and the couplings, $J_i$, $i=1,2,\dots,\ell$ between the two endpoint sites as:
\be
\varepsilon = \frac{h_1 h_2 \dots h_{\ell}}{J_1 J_2 \dots  J_{\ell-1}}\;.
\ee
Consequently, $[\ln(\varepsilon)]_{\rm av}(L) \sim \ell \sim \ln L$ and a similar $\ln L$ dependence is expected for the average mass and the average radius of gyration of the clusters. To check this assumption, we have plotted in Fig.\ref{fig:1D_parameters} the average cluster parameters, which indeed seem to follow an asymptotic $\ln L$ dependence:
\begin{align}
[\ell]_{\rm av}(L) \simeq \lambda \ln L + c_1,\nonumber\\
[\mu]_{\rm av}(L) \simeq \sigma \ln L + c_2,\nonumber\\
[R_g]_{\rm av}(L) \simeq \rho \ln L + c_3,
\label{parameters}
\end{align}
%
 
These results indicate that the energy clusters in 1D are presumably compact, i.e.~they have finite density of sites: $\lim_{L \to \infty} [\mu]_{\rm av}(L)/[\ell]_{\rm av}(L)\simeq \sigma/\lambda >0$.

\subsection{Analysis of the 2D and 3D results}

\subsubsection{Diluted model}

\begin{figure}[h!]
\begin{center}
\includegraphics[width=1. \linewidth]{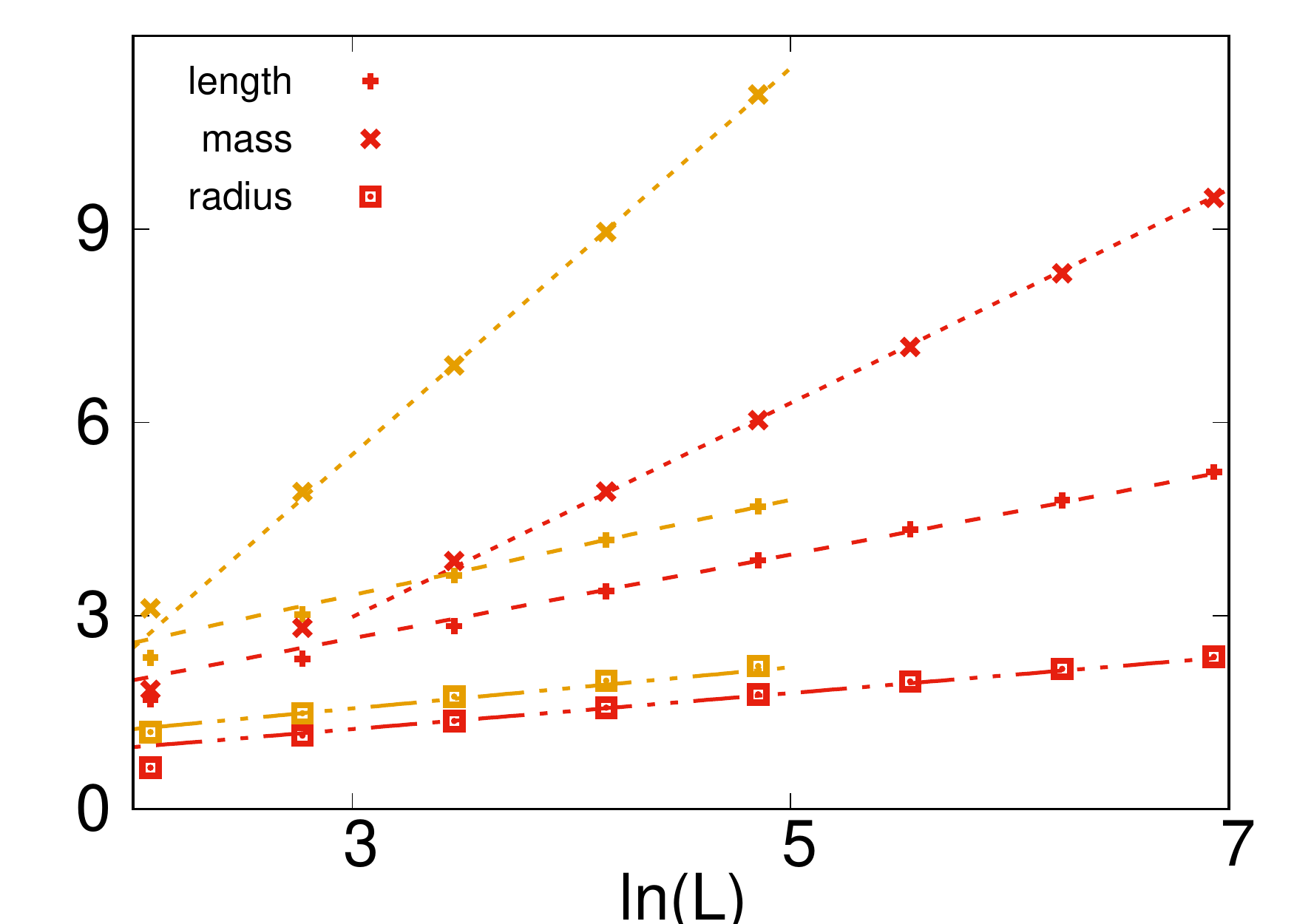}
\end{center}
\vskip-2mm
\caption{\label{fig:23D_percolation}(Color online) Average parameters of energy clusters in the 2D site-diluted TIM at $\Delta p=-0.5$, having $p=0.092746$ (red symbols) and in the 3D model at $\Delta p=-0.25$, having $p=0.0616$ (yellow symbols) as a function of the logarithm of the size of the system: the linear extension $[\ell]_{\rm av}$ with asymptotic slopes $\lambda \approx 1.66$ (2D) and $\lambda \approx 3.$ (3D); the mass $[\mu]_{\rm av}$ with asymptotic slopes $\sigma \approx 0.65$ (2D) and $\sigma \approx 0.74$ (3D) and the radius of gyration $[R_g]_{\rm av}$ with asymptotic slopes $\rho \approx 0.28$ (2D) and $\rho \approx 0.32$ (3D).}
\end{figure}

In higher dimensions, the energy clusters in the diluted TIM are equivalent to percolation clusters and their
cluster-mass distribution is known from percolation theory\cite{stauffer_aharony} to be exponential $P_D(\mu) \sim \exp[-\alpha(p) \mu]$, for large $\mu$. In a finite system of length $L$, a cluster can be placed at $\sim L^d$ different positions and the typical mass of the largest cluster $\mu_{\mathrm{typ}}$ should satisfy the relation: $P_D(\mu_\mathrm{typ})L^d \approx 1$. From this follows that $\mu_{\mathrm{typ}}$ should scale with $\ln L$. To verify this conjecture, we numerically calculated the size-dependence of various properties of the largest cluster. The average value of the mass, length and radius of gyration are shown in Fig.\ref{fig:23D_percolation} for different finite systems both in 2D and 3D.

As seen in this figure, the expected $\ln L$ scaling works very well for the average mass in 2D, as well as in 3D. More surprisingly, also the average length and the average radius of gyration seem to follow the same $\ln L$ scaling. For compact energy clusters, a relation $[\mu]_{\rm av} \sim [\ell]_{\rm av}^d$ should work, while the observed $[\mu]_{\rm av} \sim [\ell]_{\rm av}$ scaling is valid for quasi-one-dimensional (tree-like) objects. 
\begin{figure}[h!]
\begin{center}
\includegraphics[width=1. \linewidth]{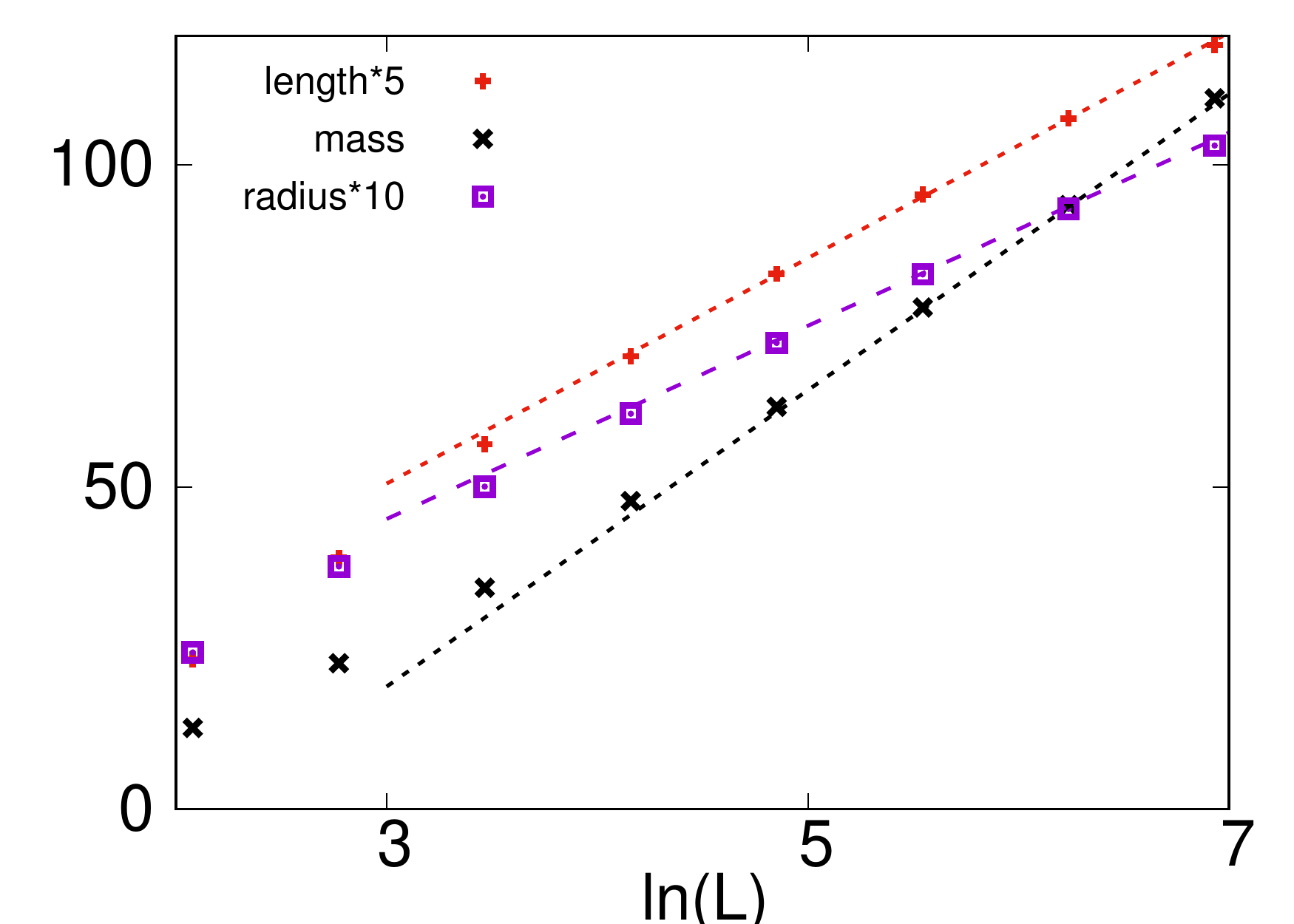}
\end{center}
\vskip-2mm
\caption{\label{fig:percolation_2D_delta02}(Color online) Average parameters of energy clusters in the 2D diluted TIM with site-percolation at $\delta=-0.2$, having $p=0.392746$ as a function of the logarithm of the size of the system: the linear extension $[\ell]_{\rm av}$ with asymptotic slope $\lambda \approx 3.5$; the mass $[\mu]_{\rm av}$ with asymptotic slope $\sigma \approx 23.$ and the radius of gyration $[R_g]_{\rm av}$ with asymptotic slope $\rho \approx 1.5$.}
\end{figure}
To verify that this result is not related to the relatively small sizes of the energy clusters we have repeated the analysis closer to the percolation threshold at $\Delta p=-0.2$ for site-percolation in 2D. The obtained results in Fig.\ref{fig:percolation_2D_delta02} show the same type of $\ln L$ scaling, although the energy clusters are comparatively large. On closer inspection of a typical energy cluster in Fig.\ref{fig:perc}, one can see an effective tree-like structure that corresponds to the observed linear mass-length relationship.

We repeated the previous analyzes for bond-diluted TIMs, where the energy clusters were selected from the extremal clusters of bond percolation and the geometric properties were qualitatively the same as for site percolation.

\subsubsection{Random model}

\begin{figure}[h!]
\begin{center}
\includegraphics[width=1. \linewidth]{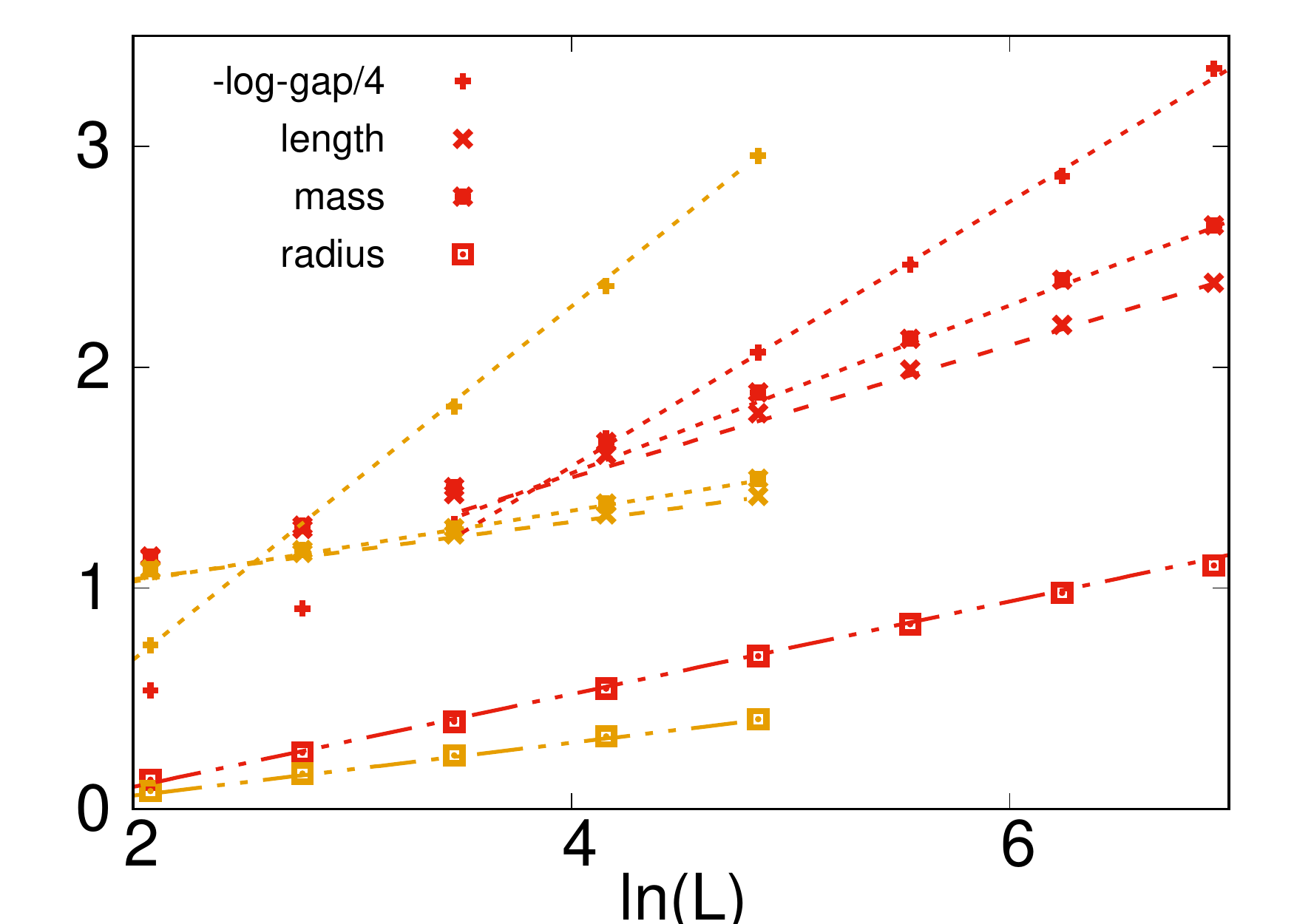}
\end{center}
\vskip-2mm
\caption{\label{fig:23D_parameters}(Color online) Average parameters of energy clusters in the random 2D RTIM at $\theta=2.6784$ (red symbols) and in the random 3D model at $\theta=3.9$ (yellow symbols) as a function of the logarithm of the size of the system. Minus log-gap $-[\ln(\varepsilon)]_{\rm av}$ with asymptotic straight lines with slopes $z \approx 2.4$ (2D) and $z \approx 3.2$; the linear extension $[\ell]_{\rm av}$ with asymptotic slopes $\lambda \approx 0.30$ (2D) and $\lambda \approx 0.13$ (3D); the mass $[\mu]_{\rm av}$ with asymptotic slopes $\sigma \approx 0.38$ (2D) and $\sigma \approx 0.16$ (3D) and the radius of gyration $[R_g]_{\rm av}$ with asymptotic slopes $\rho \approx 0.21$ (2D) and $\rho \approx 0.12$ (3D).}
\end{figure}

For the random TIMs, the cluster structure of the ground state is calculated through the numerical application of the SDRG method. In 2D, we considered the point $\theta=2.6784$ and calculated $10^6$ realizations for $L=8,16,32,64$ and $128$; $10^5$ realizations for $L=256$ and $512$ and finally $10^4$ realizations for $L=1024$. In 3D, we used $\theta=3.9$ with $10^6$ realizations for $L=8,16$ and $32$; $10^5$ realizations for $L=64$ and finally $10^4$ realizations for $L=128$. In both cases, the parameters, $[\ell]_{\rm av}(L)$, $[\mu]_{\rm av}(L)$ and $[R_g]_{\rm av}(L)$, as well as the average value of the log-gap: $[\ln(\varepsilon)]_{\rm av}(L)$ are found to have an asymptotic $\ln L$ dependence, which is shown in Fig.\ref{fig:23D_parameters}. 

Repeating the calculations at other points of the Griffiths-phase we obtained the same qualitative picture: the size-dependence of the parameters of the energy clusters are logarithmic in $L$, just the prefactors are $\theta$ dependent. Consequently, we obtained the same result as for the diluted models: the energy clusters are quasi-one-dimensional, having a tree-like topology even in 2D and 3D. We note that this observation was already made at the critical point\cite{ddRG}, which explains why the critical exponent $\psi$ in Eq. (\ref{psi}) is almost independent of dimensions. Our numerical results show that this behavior of the energy clusters seems to remain valid even in the Griffiths-phase.


\section{Discussion}
\label{sec:disc}

One peculiarity of random many-body systems is that dynamical observables can be singular not only at the critical point, but in extended regions of the paramagnetic and ferromagnetic phases as well. This singular behavior is due to rare regions, which occur with very low probability, but have a very large relaxation time, which actually diverges in the thermodynamic limit. In this paper, we studied the geometrical properties of these rare regions in the paramagnetic phase of ferromagnetic quantum magnets, such as the transverse Ising model, which are either diluted or contain random parameters (couplings and/or transverse fields). 

In diluted models, a fraction of sites or bonds are missing and the existing spins form well-defined clusters, the structure of which is described by percolation theory. On the contrary, in random models, clusters are formed during the SDRG procedure when sites become decimated when the energy-scale is lowered below a characteristic local value. In this way, clusters in diluted models are defined sharply, while in random models they depend on the energy-scale and the final cluster structure is obtained only at the lowest energy scale of the SDRG process. 

In the Griffiths-phase, these clusters are locally in the ordered phase and the rare region responsible for Griffiths singularities is associated to such cluster, which has the smallest excitation energy. In diluted models, the energy cluster also has the largest mass, while in random models it is the cluster decimated at the lowest energy scale. From a topological point of view, the energy cluster is connected for diluted models, while it contains several disconnected components for random models.

In this paper, we considered the transverse Ising model in one-, two- and three-dimensional hypercubic lattices and calculated the energy clusters in diluted and random samples of varying  lengths, $L$.
We have determined the mass (number of sites), $\mu$, the linear extension, $\ell$, and the radius of gyration, $R_g$, of the energy clusters. The average mass is found to scale as $\ln L$ for large systems.
For dilute models, this follows from extreme statistics and from the result of percolation theory that the distribution of the mass of the clusters is exponential. 

More surprisingly, the average linear parameters of the energy clusters, $[\ell]_{\rm av}$ and $[R_g]_{\rm av}$ are found to scale with $\ln L$, also in higher dimensions. Thus, from the numerical results, we observe a linear relation between mass and linear extension: $[\mu]_{\rm av} \sim [\ell]_{\rm av}$, both for diluted and random models. Regarding diluted models, the typical (not extreme) percolation clusters in the subcritical region, $p<p_c$, have a finite extent, $\sim \xi$, and these are compact and isotropic. One could na{\"i}vely expect that the same properties hold for the extreme clusters, too, which is not the case, however. As seen in  Fig.\ref{fig:perc}, the extreme clusters have a large linear extent, $\ell \gg \xi$, are isotropic, but not compact. As seen in Fig.\ref{fig:perc}, the extreme energy clusters have a tree-like structure from building blocks of the size $\xi$.

In higher dimensional random models, the energy clusters are not compact either. Previous studies performed at the critical point showed that the energy clusters are worm-like, quasi-one-dimensional\cite{ddRG}, disconnected objects. This is in agreement with the observation that the critical exponent $\psi$, defined in Eq.(\ref{psi}) is close to $0.5$ in any dimensions. Our present investigations revealed that the quasi-one-dimensional form of the energy clusters is likely in the Griffiths-phase, too. 

In our numerical work, we were deep in the paramagnetic phase, so that the relation $L \gg \xi$ is well satisfied. Being closer to the critical point, a cross-over from critical dynamics to Griffiths dynamics should take place, which is accompanied with the change of the structure and symmetry of the energy cluster.

Our results about the geometric properties of the energy cluster are expected to be valid for other models having infinite disorder scaling and strong Griffiths effects. For diluted models, the cluster structure evidently does not depend on the actual form of the dynamical variable. For random models, the TIM fixed point is expected to govern the critical properties of models having a discrete symmetry (q-state Potts model\cite{senthil,anfray}, Ashkin-Teller model\cite{carlon-clock-at,barghathi,chatelain}, etc.), as well as random stochastic models, such as the contact process\cite{igloicontact,vojta_dickison}. For these models, the form of the Griffiths singularities is very similar, which should hold for the basic geometrical properties of the energy clusters, too\cite{clusters}. 

\begin{acknowledgments}
This work was supported by the National Research Fund under Grants No. K128989 and No. KKP-126749
F.I. is indebted to L. Gr\'an\'asy for useful discussions.
\end{acknowledgments}

\end{document}